\documentclass[runningheads,fleqn]{llncs}

\usepackage{booktabs}   
\usepackage{subcaption} 

\usepackage{latexsym}
\usepackage{amssymb}
\usepackage{amsmath}
\usepackage{stmaryrd}
\usepackage[all]{xy}
\usepackage{graphicx}
\usepackage{hyperref}
\usepackage{multicol}
\usepackage{bm}
\usepackage{cancel}

\usepackage{tikz}
\usetikzlibrary{calc,arrows,shadows,positioning}

\newcommand{\true}{\mathit{true}}
\newcommand{\false}{\mathit{false}}

\newcommand{\sched}{\mathsf{sched}}

\newcommand{\ignore}[1]{}

\usepackage{xcolor}

\newcommand{\red}[1]{{\color{red} #1}}

\newcommand{\nil}{\epsilon}

\newcommand{\trace}{\mathit{tr}}

\newcommand{\cons}{\!:\!}
\newcommand{\conc}{\mbox{$;$}}

\long\def\comment#1{}

\newcommand{\cauder}{\textsf{CauDEr}}

\newcommand{\spawn}{\mathsf{spawn}}
\newcommand{\send}{\mathsf{send}}
\newcommand{\deliver}{\mathsf{deliver}}
\newcommand{\rec}{\mathsf{rec}}

\newcommand{\race}{\mathsf{race\_set}}
\newcommand{\rdep}{\mathsf{rdep}}
\newcommand{\variant}{\mathsf{variant}}
\newcommand{\match}{\mathsf{match}}
\newcommand{\actions}{\mathsf{actions}}

\usepackage{accents}

\makeatletter
\providecommand{\leftsquigarrow}{%
  \mathrel{\mathpalette\reflect@squig\relax}%
}
\newcommand{\reflect@squig}[2]{%
  \reflectbox{$\m@th#1\rightsquigarrow$}%
}
\makeatother

\begin{document}

\title{%
Computing Race Variants in\\ Message-Passing 
Concurrent Programming\\
with Selective Receives%
\thanks{
This work has been partially supported by 
grant PID2019-104735RB-C41
funded by MCIN/AEI/ 10.13039/501100011033, by  
\emph{Generalitat Valenciana} under grant Prometeo/2019/098 
(DeepTrust), 
and by French ANR project DCore ANR-18-CE25-0007.
}
}

\titlerunning{Computing Race Variants in Message-Passing 
Concurrent Programming}

\author{Germ\'an Vidal\orcidID{0000-0002-1857-6951}}

\authorrunning{G. Vidal}

\institute{
  VRAIN, Universitat Polit\`ecnica de Val\`encia\\
\email{gvidal@dsic.upv.es}
}

\maketitle

\begin{abstract}
Message-passing concurrency is a popular computation model that 
underlies several programming languages like, e.g., Erlang, Akka, 
and (to some extent) Go and Rust. 
In particular, we consider a message-passing concurrent
language with dynamic process spawning and
\emph{selective} receives, i.e., where messages 
can only be consumed by the target process when they match 
a specific constraint (e.g., the case of Erlang). 
In this work, we introduce a  notion of \emph{trace} that 
can be seen as an abstraction of a class of 
\emph{causally equivalent} executions (i.e., which produce
the same outcome). We then show that 
execution traces can be used to identify message 
\emph{races}. We provide constructive definitions to 
compute message races as well as to produce so-called
\emph{race variants}, which can then 
be used to drive new executions
which are not causally equivalent to the previous ones.
This is 
an essential ingredient of state-space exploration techniques for 
program verification.\\[2ex]
  Published as \emph{Vidal, G. (2022). Computing Race Variants in Message-Passing Concurrent Programming with Selective Receives. In: Mousavi, M.R., Philippou, A. (eds) Formal Techniques for Distributed Objects, Components, and Systems. FORTE 2022. Lecture Notes in Computer Science, vol 13273. Springer, Cham.}
\\[2ex]
The final authenticated publication is available online at\\
\url{https://doi.org/10.1007/978-3-031-08679-3\_12}
\end{abstract}


\section{Introduction} \label{sec:intro}

Software verification and debugging are recognized as 
essential tasks in the field of software development. 
Not surprisingly, a recent study \cite{UndoReport} points out that 
26\% of developer time is spent reproducing and fixing code bugs
(which adds up to \$61 billion annually). The study also
identifies \emph{reproducibility} of bugs as the biggest
challenge to fix bugs faster.
The situation is especially difficult for concurrent and 
distributed applications because of nondeterminism.
In this context, traditional testing techniques often provide
only a poor guarantee regarding software correctness.

As an alternative, \emph{state-space exploration} techniques 
constitute a well established approach
to the verification of concurrent software 
that basically consists in exploring
the reachable states of a program, checking whether a given
property holds (like some type of deadlock, a runtime error, etc).
This is the case, e.g., of \emph{model checking} \cite{CES86},
where properties have been traditionally verified using a 
\emph{model} of the program. More recently, several
\emph{dynamic} approaches to state-space exploration
have been introduced, which work directly with
the implementation of a program.
\emph{Stateless model checking} \cite{God97} and 
\emph{reachability testing}
\cite{Tai97,LT02} are examples of this approach.
In turn, reproducibility of bugs has been tackled
by so-called \emph{record-and-replay} debuggers. In this case,
a program is first instrumented so that its execution produces
a \emph{log} as a side-effect. If a problem occurs during
the execution of the program, one can use the generated
log to play it back in the debugger and try to locate the
source of the misbehavior.

In this work, we focus on an asynchronous 
\emph{message-passing} concurrent
programming language like, e.g., Erlang \cite{erlang},
Akka \cite{akka} and, to some extent, Go \cite{golang}
and Rust \cite{rust}.
A running application consists of a number of
processes, each with an associated (private) \emph{mailbox}.
Here, 
processes can only interact through (asynchronous) 
message sending and receiving,
i.e., we do not consider shared-memory operations.
Typically, there is some degree of nondeterminism in
concurrent executions that may
affect the outcome of a computation. For instance, when two
processes send messages to another process,
these messages may sometimes arrive in any order. 
These so-called 
\emph{message races} play a key role in the execution of
message-passing concurrent programs, and exploring
all feasible combinations of message sending and
receiving is an essential component of state-space exploration techniques.

In particular, we consider a language
with so-called \emph{selective} receives, where a process does
not necessarily consume the messages in its mailbox 
in the same order they were delivered, since receive statements
may impose additional constraints. 
For instance, a receive statement in Erlang has the form
\[
\texttt{receive $p_1$ [when $g_1$] $\to t_1$; \ldots
; $p_n$ [when $g_n$] $\to t_n$ end}
\]
In order to evaluate this statement, a process should
look for the \emph{oldest} message in its mailbox that
matches a pattern $p_i$ and the corresponding
(optional) guard $g_i$ holds 
(if any);\footnote{If the message matches 
several patterns, the first one is considered.} 
in this case, the process
continues with the evaluation of expression $t_i$. 
When no message matches any pattern, the execution of the
process is \emph{blocked} until a matching message reaches
its mailbox. 

Considering a message-passing concurrent language with 
selective receives is relevant in order to deal with a language 
like Erlang. Unfortunately, current approaches either
do not consider selective receives---the case of 
reachability testing \cite{LC06} which, in contrast, considers
different ports for receive statements---or 
have not formally defined the semantics of the language and
its associated \emph{happened-before} relation---the
case of Concuerror \cite{CGS13}, which implements a
stateless model checker for Erlang that follows the approach
in \cite{AAJS17,AJLS18,AS17}.

In this paper, we introduce a notion of \emph{trace} that is tailored
to message-passing execution with dynamic process
spawning and selective 
receives. Our traces can be seen as an extension of
the \emph{logs} of
\cite{LPV19,LPV21}, which were introduced in the
context of causal-consistent \emph{replay} (reversible) 
debugging in Erlang.
In particular, the key extension consists in adding 
some additional information to the events of a trace, namely
the identifier of the target process
of a message and its actual value for \emph{send} events,
and the actual constraints of a receive statement
for \emph{receive} events. 
In this way, we can identify
not only the communications performed in
a program execution but also
its message races (see the discussion in the next section).

In contrast to other notions of trace that represent a
particular interleaving, our traces
(analogously to the \emph{logs} of \cite{LPV19,LPV21}
and the \emph{SYN-sequences}  of \cite{LC06})
record the sequence of actions performed
by each process in an execution, ignoring the concrete
scheduling of all processes' actions. These traces 
can easily be obtained by instrumenting the source code
so that each process keeps a record of its own actions.
The traces can be seen 
as an abstraction of a class of executions which are 
\emph{causally equivalent}. Roughly
speaking, two executions are causally equivalent when 
the executed actions and the final outcome are the same
but the particular scheduling might differ. 
We then introduce constructive definitions for computing
message races and \emph{race variants} from a given trace. 
Here, race variants are denoted by a (possibly partial) trace
which can then be used to drive a new program execution
(as in the replay debugger \cauder\ \cite{cauder}). 
Moreover, we prove that any execution that follows
(and possibly goes beyond) the computed
race variant cannot give rise to an execution which is 
causally equivalent to the previous one, 
an essential property of state-space exploration
techniques.

The paper is organized as follows. After some motivation 
in Section~\ref{sec:motivation}, we formalize the notions
of \emph{interleaving} and \emph{trace} 
in Section~\ref{sec:model}, where we also provide a
declarative definition of message race and prove a
number of properties.
Then, Section~\ref{sec:races} provides constructive
definitions for computing message races and race variants,
and proves that race variants indeed give rise to
executions which are not causally equivalent to the
previous one. Finally, Section~\ref{sec:relwork}
presents some related work and concludes.

\section{Message Races and Selective Receives} \label{sec:motivation}

In this section, we informally introduce the considered setting
and motivate our definition of  \emph{trace}.
As mentioned before, we consider a message-passing 
(asynchronous) concurrent language with selective receives.
Essentially, concurrency follows the
actor model: at runtime, an application can be seen
as a collection of processes that interact through message 
sending and receiving. Each process has an associated
identifier, called \emph{pid} (which stands for process identifier), 
that is unique in the execution.\footnote{In the following, 
we often say ``process $p$" to mean
``process with pid $p$".}
Furthermore, we assume that processes can be spawned 
\emph{dynamically} at runtime. 

As in other techniques where message races are computed, 
e.g., \emph{dynamic partial order reduction} (DPOR)
\cite{FG05,AAJS17} for \emph{stateless model checking} \cite{God97},
we distinguish \emph{local} evaluations from
\emph{global} (or \emph{visible}) actions. 
Examples of local evaluations are, e.g., a function
call or the evaluation of a case expression. In turn, 
\emph{global} actions include the spawning of a 
new process as well as any event
related with message passing.
In particular, we consider that sent messages are
eventually stored in the \emph{mailbox} of the target process. 
Then, the target process can
\emph{consume} these messages using a \emph{receive}
statement, which we assume is selective, i.e., 
it may impose some additional constraints on the 
receiving messages.

\entrymodifiers={++[o][F-]}
\SelectTips{cm}{}

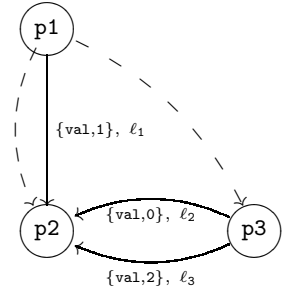
\begin{figure}[t]
\begin{minipage}{.68\linewidth}
\begin{verbatim}
  proc1() -> P2 = spawn(proc2()),
             P3 = spawn(proc3(P2)),
             send({val,1},P2).
  proc2() -> receive
                {val,M} when M>0 -> {ok,M};
                error -> error
             end.
  proc3(P2) -> send({val,0},P2),
               send({val,2},P2).
\end{verbatim}
\end{minipage}
\begin{minipage}{.3\linewidth}
$
\xymatrix@C=2cm@R=2cm{
\mathtt{p1} \ar[d]^{\mathtt{\{val,1\},~\ell_1}}
\ar@{-->}@/_1pc/[d]
\ar@{-->}@/^1pc/[dr]\\
\mathtt{p2} &
\mathtt{p3} \ar@/^1pc/[l]^{\mathtt{\{val,2\},~\ell_3}} 
\ar@/_1pc/[l]^{\mathtt{\{val,0\},~\ell_2}}
}
$
\end{minipage}
\caption{A simple message-passing program}\label{fig:example}
\end{figure}

\begin{figure}[t]
\begin{minipage}{.46\linewidth}
\centering
\fbox{\begin{tikzpicture}
\draw[->,dashed,thick] (0,1.6) node[above]{\sf p1} -- (0,0);
\draw[->,dashed,thick] (2,1.6) node[above]{\sf p2} -- (2,0);
\draw[->,dashed,thick] (4,1.6) node[above]{\sf p3} -- (4,0);

\draw[->, thick] (0,1.5) node[left]{$s_1$} -- (2,1.3) node[midway,above]{$\ell_1$} node[right]{$r_1$};

\draw[->, dotted] (4,1.1) node[right]{$s_2$} -- (2,0.9) node[midway,above]{$\ell_2$} node[left]{~};

\draw[->, dotted] (4,0.6) node[right]{$s_3$} -- (2,0.4) node[midway,above]{$\ell_3$} node[left]{~};
\end{tikzpicture}}\\[1ex]
(a)
\end{minipage}
~
\begin{minipage}{.46\linewidth}
\centering
\fbox{\begin{tikzpicture}
\draw[->,dashed,thick] (0,1.6) node[above]{\sf p1} -- (0,0);
\draw[->,dashed,thick] (2,1.6) node[above]{\sf p2} -- (2,0);
\draw[->,dashed,thick] (4,1.6) node[above]{\sf p3} -- (4,0);

\draw[->, thick] (0,1.1) node[left]{$s_1$} -- (2,0.9) node[midway,above]{$\ell_1$} node[right]{$r_1$};

\draw[->, dotted] (4,1.5) node[right]{$s_2$} -- (2,1.3) node[midway,above]{$\ell_2$} node[left]{~};

\draw[->, dotted] (4,0.6) node[right]{$s_3$} -- (2,0.4) node[midway,above]{$\ell_3$} node[left]{~};
\end{tikzpicture}}\\[1ex]
(b)
\end{minipage}
\caption{Alternative interleavings for the execution of the program
of Figure~\ref{fig:example}. 
We have three processes, identified  by  pids 
$\mathsf{p1}$, $\mathsf{p2}$
and $\mathsf{p3}$. Solid arrows denote the connection between
messages sent and received (similarly to the synchronization 
pairs of \cite{LC06}), while dotted arrows represent messages
sent but not yet received. Time, represented
by dashed lines, flows from top to bottom.} \label{fig:problema}
\end{figure}
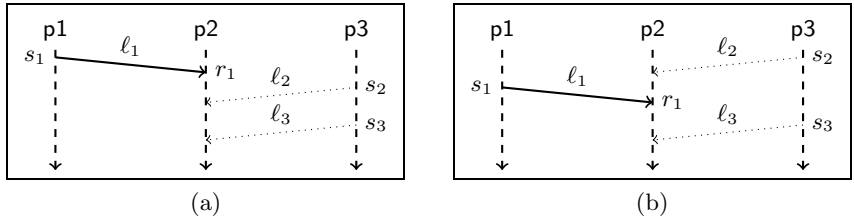

\begin{example} \label{ex:intro}
Let us consider the simple code 
shown in Figure~\ref{fig:example}
(we use a pseudocode
that resembles the syntax of Erlang \cite{erlang}).
Assume that the initial process (the one that starts the execution)
has  pid \verb$p1$ and 
that it begins with a call to function \verb$proc1$, which
first spawns two new processes with pids
\verb$p2$ and \verb$p3$ that will evaluate the calls
\verb$proc2()$ and \verb$proc3(P2)$,
respectively. A call to \verb$spawn$ 
returns the pid of the new process, so
variables \verb$P2$ and \verb$P3$ are bound to 
pids \verb$p2$ and \verb$p3$, respectively.
The evaluation of \verb$proc1()$ ends by sending
the message  \verb${val,1}$ to process 
\verb$p2$. Process spawning is denoted by a
dashed arrow in the diagram, while message sending 
is denoted by a solid arrow. Messages are \emph{tagged}
with a unique identifier (e.g., $\ell_1$).

Process \verb$p3$ sends two messages,
\verb${val,0}$ and \verb${val,2}$, to process \verb$p2$.
Process \verb$p2$ initially blocks waiting for the arrival
of a message that matches either the pattern 
\verb${val,M}$, i.e., a tuple whose first component is
the constant \verb$val$ and the second component (denoted
by variable \verb$M$) is an integer greater than zero, or the
constant \verb$error$. Note that message
$\tt \{val,0\}$ does not match the constraints of the receive
statement since the integer value is not greater than zero.
Thus, only messages $\ell_1$ and $\ell_3$ can be
consumed by the receive statement of process \verb$p2$.
\end{example}
In principle, one could represent a program execution 
by means of a concrete \emph{interleaving} of its
concurrent actions, i.e., a sequence of 
\emph{events} of the form $pid\cons action$. E.g.,
we could have the following interleaving for the program
of Figure~\ref{fig:example}: 
\begin{verbatim}
  (1) p1:spawn(proc2())           (4) p2:receive({val,1})
  (2) p1:spawn(proc3(p2))         (5) p3:send({val,0},p2)   
  (3) p1:send({val,1},p2)         (6) p3:send({val,2},p2)
\end{verbatim}
This interleaving is
graphically depicted in Figure~\ref{fig:problema}a,
where process spawning is omitted for clarity.

In this work, though, 
we opt for a different representation,
which is similar to the notion of \emph{log} (in the context of
replay debugging \cite{LPV19,LPV21}) and that of
\emph{SYN-sequence} (in reachability testing \cite{LC06}).
In contrast to  interleavings
(as in, e.g., stateless model checking  \cite{God97}
and DPOR techniques \cite{FG05,AAJS17}), the advantage of
using logs 
is that they
represent a \emph{partial order} 
for the concurrent actions, so that DPOR
techniques are no longer needed.
To be more precise, 
a log (as defined in \cite{LPV19}) 
maps each process to
a sequence of the following actions:
\begin{itemize}
\item process spawning, denoted by $\spawn(p)$, where $p$
is the \emph{pid} of the new process;
\item message sending, denoted by $\send(\ell)$, 
where $\ell$ is a  message tag;
\item and message reception, denoted by $\rec(\ell)$, 
where $\ell$ is a message tag.
\end{itemize}
In contrast to the SYN-sequences of  \cite{LC06}, 
\emph{synchronization pairs}
(connecting message sending and receiving)
are not explicitly considered but can easily be inferred 
from send/receive actions with the same message tag.
Furthermore, logs include $\spawn$ actions because  
runtime processes are not statically fixed, which is
not considered by SYN-sequences.
Logs are used by the 
reversible debugger
\cauder\ \cite{cauder} as part of an approach to 
\emph{record-and-replay} debugging in Erlang 
(a popular approach to deal with the problem of reproducibility 
of bugs).

In practice, logs can be obtained by using an instrumented
semantics (as in \cite{LPV19,LPV21}) or by instrumenting
the program so that its execution (in the standard 
environment) produces a log as a side-effect
(along the lines of the technique presented in \cite{GV21lopstr}). 
It is worthwhile to note that no centralised monitoring is required; 
every process only needs to register its own actions independently. 
For instance, a log associated with the execution shown
in Figure~\ref{fig:example} is as follows:
\[
[\mathsf{p1}\mapsto \spawn(\mathsf{p2}),\spawn(\mathsf{p3}),\send(\ell_1);~~
\mathsf{p2}\mapsto  \rec(\ell_1);~~
\mathsf{p3}\mapsto \send(\ell_2),\send(\ell_3)]
\]
Unfortunately, this log does not contain enough information
for computing \emph{message races}.
A first, obvious problem is that \emph{send} events do not include
the pid of the target process. Hence, even if there is a
(potential) race between messages $\ell_1$ and $\ell_2$ to reach
process $\mathsf{p2}$, this cannot
be determined from the log. Trivially, one could
solve this problem by adding the pid of the target process
to every $\send$ event, as follows (we omit $\spawn$ 
actions since they are not relevant for the discussion):
\[ \label{eqn:trace}
[\mathsf{p1}\mapsto 
\ldots,\send(\ell_1,\mathsf{p2});~~
\mathsf{p2}\mapsto  \rec(\ell_1);~~
\mathsf{p3}\mapsto \send(\ell_2,\mathsf{p2}),\send(\ell_3,\mathsf{p2})]
\tag{*}
\]
Now, 
in principle, one could say that messages $\ell_1$ and $\ell_2$ race
for process \verb$p2$ 
since the target is the same ($\mathsf{p2}$) and there are no
dependencies among  
$\send(\ell_1,\mathsf{p2})$ and 
$\send(\ell_2,\mathsf{p2})$
($s_1$ and $s_2$ in Figure~\ref{fig:problema}a).

However, when we consider \emph{selective} receives, the log
\eqref{eqn:trace} above can be ambiguous. In particular, this 
log represents both interleavings represented by the
diagrams of Figure~\ref{fig:problema}a and 
Figure~\ref{fig:problema}b
(where message $\ell_2$ 
reaches first process $\mathsf{p2}$ but its associated
value, $\tt \{val,0\}$, does not match the constraints of the
the receive statement since the guard \verb$M>0$ does not hold
for \verb$M=0$). However, the diagram in 
Figure~\ref{fig:problema}a points out to a (potential) message
race between messages $\ell_1$ and $\ell_2$, while the
diagram in Figure~\ref{fig:problema}b 
suggests a (potential) message race between messages
$\ell_1$ and $\ell_3$ instead.


In order to distinguish the executions shown in Figures~\ref{fig:problema}a
and \ref{fig:problema}b one could  add  a new action,
$\deliver(\ell)$, to explicitly account for the delivery of a
message with tag $\ell$. In this way, we would 
know the order in which messages
are stored in the process' mailbox, which uniquely determines
the order in which they can be consumed by receive statements.
E.g., the actions of process $\mathsf{p2}$ in the execution of
Figure~\ref{fig:problema}a would be
\[
\label{eqn:trace2}
%
[~\ldots~\mathsf{p2} \mapsto  
\deliver(\ell_1),\rec(\ell_1),\deliver(\ell_2),\deliver(\ell_3) 
~\ldots ~]
\tag{**}
\]
while those of Figure~\ref{fig:problema}b  would be as follows:
\[
\label{eqn:trace3}
[~\ldots~\mathsf{p2}\mapsto  
\deliver(\ell_2),\deliver(\ell_1),\rec(\ell_1),\deliver(\ell_3) 
~\ldots~]
\tag{***}
\]
This approach is explored in \cite{GV21arxiv}, where a 
\emph{lightweight} (but approximate)
technique to computing message races is proposed.
Unfortunately, making explicit message delivery
does not suffice to allow one to compute message
races in general.
In particular, while it would allow us to distinguish
the situation of Figure~\ref{fig:problema}a from that
of Figure~\ref{fig:problema}b, we could not
still determine whether there is an actual race between
messages $\ell_1$ and $\ell_2$ 
or between messages $\ell_1$ and $\ell_3$.
For instance, for the program of Figure~\ref{fig:example}, 
only the message race between $\ell_1$ and $\ell_3$ is
feasible, as explained above.
For this purpose, we need to also include 
the actual values of messages
as well as the constraints of receive statements.

In the next section, we propose an appropriate 
definition of \emph{trace}
(an extended log) that includes enough information for
computing message races.

\section{Execution Traces} \label{sec:model}

In this section, we formalize an appropriate notion of
trace that is adequate to compute message races
in a message-passing concurrent language with 
dynamic process spawning and selective receives.
Here, we do not consider a specific programming language
but formalize our developments in the context of a
generic language that includes the basic actions
$\spawn$, $\send$, and $\mathsf{receive}$.

As mentioned in the previous section, we consider that
each process is uniquely identified by a pid.
A message takes a \emph{value} $v$ (from a given domain) 
and is uniquely identified by a tag 
$\ell$.\footnote{Message tags were  
introduced in \cite{LNPV18jlamp}
to uniquely identify messages, since we might have
several messages with the same value and would be
indistinguishable otherwise.
} 
We further require the domains of pids, values, and
tags to be disjoint.
We also consider a generic domain 
of constraints and a decidable function $\match$
so that, for all value $v$ and constraint $cs$,
$\match(v,cs)$ returns $\mathit{true}$ if the value matches
the constraint $cs$ and $\false$ otherwise.
In Erlang, for instance, a constraint is associated with
the clauses of a receive statement, i.e., 
it has the form
\texttt{$(p_1\!\!$ [when $g_1$] $\!\!\to t_1$; \ldots
;$p_n\!\!$ [when $g_n$] $\!\!\to t_n)$},
and function $\match$
determines if a value $v$ matches some pattern $p_i$
and the associated guard $g_i$ (if any)
evaluates to true.

In this work, events have the form $p\cons a$, where $p$ is
a pid and $a$ is one of the following actions:
\begin{itemize}
\item $\spawn(p')$, which denotes the spawning of
a new process with pid $p'$;
\item $\send(\ell,v,p')$, which denotes the sending of a
message with tag $\ell$ and value $v$ to process $p'$;
\item $\rec(\ell,cs)$, which denotes the \emph{reception} 
of a message with tag $\ell$ by a receive statement with 
a constraint $cs$.\footnote{Note that \emph{receive} actions
represent the \emph{consumption} of messages 
by receive statements 
rather than their delivery to the
process' mailbox. Observe that the
order of message delivery and message 
reception might be different
(see, e.g., messages $\ell_1$ and $\ell_2$ in
Figure~\ref{fig:problema}b).}
\end{itemize}
In the following, a (finite) sequence is denoted as
follows: $e_1,e_2,\ldots,e_n$, $n\geq 0$,
where $n$ is the length of the sequence. 
We often use
set notation for sequences and let $e\in S$  denote
that event $e$ occurs in sequence $S$.
Here, $\nil$ denotes an empty sequence, while
$S\conc S'$ denotes the concatenation of 
sequences $S$ and $S'$; by abuse of notation, 
we use the same operator  when a sequence has only
a single element, i.e., $e_1\conc (e_2,\ldots,e_n)$ and
$(e_1,\ldots, e_{n-1})\conc e_n$ both denote the 
sequence $e_1,\ldots,e_n$.
Furthermore, given a sequence of events 
\[
S = (p_1\cons a_1,p_2\cons a_2,\ldots,p_n\cons a_n)
\]
we let
$\actions(p,S)$ denote the sequence of actions
$
(a'_1,a'_2,\ldots,a'_m)
$ 
such that $p\cons a'_1,p\cons a'_2,\ldots,
p\cons a'_m$ are all and only 
the events of process $p$ in $S$ and in the
same order.
Given a sequence $S=(e_1,\ldots,e_n)$, we also
say that $e_i$ \emph{precedes} $e_j$, in symbols
$e_i\prec_S e_j$, if $i<j$.

Now, we formalize the notions of \emph{interleaving}
and \emph{trace}. Intuitively speaking, an interleaving is a
sequence of events that represents a \emph{linearization} of the 
actions of a concurrent execution, while a trace is a mapping
from processes to sequences of actions (so a trace only
denotes a \emph{partial relation} on events).

\begin{definition}[interleaving] \label{def:interleaving}
  A sequence of events $S=(p_1\cons a_1,\ldots,
  p_n\cons a_n)$ is an \emph{interleaving} with initial
  pid $p_1$ if the following conditions hold:
  \begin{enumerate}
  \item Each event $(\red{p_j}\cons a_j)\in S$ is either 
  preceded  by an event $(p_i\cons\spawn(\red{p_j}))\in S$ 
  with $p_i\neq p_j$,   $1\leq i < j \leq n$, or $p_j=p_1$.
  \item Each event $(p_j\cons\rec(\red{\ell},cs))\in S$ is preceded by
  an event $(p_i\cons \send(\red{\ell},v,p_j))\in S$, $1\leq i < j \leq n$,
  such that $\match(v,cs)=\true$.
  \item For each pair of events 
  $p_i\cons\send(\ell,v,p_j),~p_j\cons \rec(\ell,cs)\in S$, 
  we have that, for all $p_i\cons\send(\ell',v',p_j)\in S$ 
  that precedes $p_i\cons\send(\ell,v,p_j)$, in symbols
  \[
  p_i\cons\send(\ell',v',p_j)
  \prec_S
  p_i\cons\send(\ell,v,p_j)
  \]
  either $\match(v',cs)=\false$ or there is an 
  event $p_j\cons\rec(\ell',cs')\in S$ such that
  $
    p_j\cons\rec(\ell',cs')
    \prec_S
   p_j\cons \rec(\ell,cs)
   $.
  \item Finally, for all event $p_i\cons\spawn(p_j)$, 
  $p_j$ only occurs as the argument of $\spawn$ in this
  event, and for all event $p_i\cons\send(\ell,v,p_j)$, 
  $\ell$ only occurs as the argument of $\send$ in this
  event (uniqueness of pids and tags).

  \end{enumerate}
\end{definition}
The first two conditions in the definition of interleaving are
very intuitive: all the actions of a process (except for those
of the initial process, $p_1$)
must happen after its spawning, and each reception of a 
message $\ell$ must be preceded by a sending of 
message $\ell$ and, moreover, the message value should 
match the receive constraint. 
The third condition is a bit more involved
but can be explained as follows: the messages sent between 
two given processes should be delivered in the same order 
they were sent. Thus, if a process $p_j$ receives a message
$\ell$ from process $p_i$, all previous messages sent 
from $p_i$ to $p_j$ (if any) should have been already
received or their value should not match the constraint of
the receive statement.
The last condition
simply ensures that pids and tags are unique in an
interleaving, as mentioned before.

\begin{example} \label{ex:interleaving}
  Consider the program of Example~\ref{fig:example}.
  A possible interleaving is 
  \[ 
  \begin{array}{l}
  \mathsf{p1}\cons\spawn(\mathsf{p2}),~
  \mathsf{p1}\cons\spawn(\mathsf{p3}),~
  \mathsf{p1}\cons\send(\ell_1,v_1,\mathsf{p2}),~
  \mathsf{p2}\cons\rec(\ell_1,cs_1),\\
  \mathsf{p3}\cons\send(\ell_2,v_2,\mathsf{p2}),~
  \mathsf{p3}\cons\send(\ell_3,v_3,\mathsf{p2})\\
  \end{array}
  \]
  which can be graphically represented by the diagram
  of Figure~\ref{fig:problema}a. 
\end{example}
An interleaving induces a \emph{happened-before}
relation \cite{Lam78} on events as follows:

\begin{definition}[happened-before, independence]
\label{def:happened-before}
  Let $S=(e_1,\ldots,e_n)$ be an interleaving. We say that 
  $e_i=(p_i\cons a_i)$ happened before $e_j=(p_j\cons a_j)$,
  $i<j$, in symbols $e_i \leadsto_S e_j$, if one of the following  
  conditions hold:
  \begin{enumerate}
  \item $p_i=p_j$ (i.e., the actions of a given process
  cannot be swapped);
  \item $a_i = \spawn(p_j)$ (i.e., a process cannot perform an
  action before it is spawned);
  \item $a_i = \send(\ell,v,p_j)$ and $a_j = \rec(\ell,cs)$ (i.e.,
  a message cannot be received before it is sent).
  \end{enumerate}
  If $e_i\leadsto_S e_j$ and $e_j\leadsto_S e_k$, then 
  $e_i\leadsto_S e_k$
  (transitivity). If neither $e_i\leadsto_S e_j$ nor 
  $e_j \leadsto_S e_i$,
  we say that the two events are independent.
\end{definition}
Given an interleaving $S$, the associated
happened-before relation $\leadsto_S$ is clearly
a (strict) partial order since the following properties hold:
\begin{itemize}
\item No event may happen before itself (irreflexivity)
since $e_i \leadsto_S e_j$ requires $i<j$ by
definition.\footnote{Note that repeated events in an interleaving
are not allowed by Definition~\ref{def:interleaving}.}

\item If $e_i \leadsto_S e_j$ we have $i<j$ and, thus,
$e_j \leadsto_S e_i$ is not possible (asymmetry). 

\item Finally, the relation $\leadsto_S$ is transitive
by definition.
\end{itemize}
In the following, we say that two interleavings are 
\emph{causally equivalent} if they have the same
events and only differ in the swapping of a number of
independent events. Formally,

\begin{definition}[causal equivalence] 
\label{def:causal-equivalence}
  Let $S_1$ and $S_2$ be interleavings with the
  same initial pid. We say that $S_1$ 
  and $S_2$ are \emph{causally equivalent}, in symbols 
  $S_1\approx S_2$, if $S_2$ can be
  obtained from $S_1$ by a finite number of swaps 
  of consecutive independent events.
\end{definition}
We note that our notion of causal equivalence 
is similar to that of \emph{trace equivalence} in 
\cite{Maz86} and that of causally equivalent 
\emph{derivations} in \cite{LPV19,LPV21}.

The causal equivalence relation on interleavings is
an \emph{equivalence relation} since it is
trivially reflexive ($S\approx S$ holds for all interleavings), 
symmetric ($S_1\approx S_2$ implies $S_2\approx S_1$
by considering the same swaps in the reverse order),
and transitive ($S_1\approx S_2$ and $S_2\approx S_3$
implies $S_1\approx S_3$ by considering first
the swaps that produce $S_2$ from $S_1$ and, then, those that
transform $S_2$ into $S_3$).
%

It is worthwhile to note that not all independent events
can be swapped if we want to produce a valid interleaving.
Let us illustrate this point with an example:

\begin{example}
  Consider the interleaving shown in Example~\ref{ex:interleaving}
  that is graphically 
  represented in the diagram of Figure~\ref{fig:problema}a.
  Here, we might perform a number of swaps of independent
  consecutive events so that we end up with the following 
  causally equivalent interleaving:
  \[ 
  \begin{array}{l}
  \mathsf{p1}\cons\spawn(\mathsf{p2}),~
  \mathsf{p1}\cons\spawn(\mathsf{p3}),~
  \red{\mathsf{p3}\cons\send(\ell_2,v_2,\mathsf{p2})},~
  \mathsf{p1}\cons\send(\ell_1,v_1,\mathsf{p2}),\\
  \mathsf{p2}\cons\rec(\ell_1,cs_1),~
  \mathsf{p3}\cons\send(\ell_3,v_3,\mathsf{p2})\\
  \end{array}
  \]
  which corresponds to the diagram
  of Figure~\ref{fig:problema}b. 
  In this case, we were able to swap the events
  $\mathsf{p1}\cons\send(\ell_1,v_1,\mathsf{p2})$
  and
  $\mathsf{p3}\cons\send(\ell_2,v_2,\mathsf{p2})$
  because they are independent and, moreover,
  the resulting interleaving does not violate condition~(3)
  in Definition~\ref{def:interleaving} since 
  $\match(v_2,cs_1) = \false$.

  In contrast, we could not swap 
  $\mathsf{p1}\cons\send(\ell_1,v_1,\mathsf{p2})$
  and
  $\mathsf{p3}\cons\send(\ell_3,v_2,\mathsf{p2})$
  since the resulting sequence of events
  \[ 
  \begin{array}{l}
  \mathsf{p1}\cons\spawn(\mathsf{p2}),~
  \mathsf{p1}\cons\spawn(\mathsf{p3}),~
  \mathsf{p3}\cons\send(\ell_2,v_2,\mathsf{p2}),~
  \red{\mathsf{p3}\cons\send(\ell_3,v_3,\mathsf{p2})},\\
  \mathsf{p1}\cons\send(\ell_1,v_1,\mathsf{p2}),~
  \mathsf{p2}\cons\rec(\ell_1,cs_1)\\
  \end{array}
  \]
  would not be an interleaving because it would violate condition~(3)
  in Definition~\ref{def:interleaving}; namely, we have 
  $\match(v_3,cs_1)=\true$ and, thus,
  event $\mathsf{p2}\cons\rec(\ell_1,cs_1)$
  would not  be correct in this position (message $\ell_3$
  should be received instead).
\end{example}
In general, we can easily prove that the swap of two independent
events in an interleaving always produces a valid interleaving
(according to Definition~\ref{def:interleaving})
except when the considered events are both $\send$
with the same source and target pids.
In this last case, it depends on the particular interleaving,
as illustrated in the previous example.

A straightforward property is the following: 
causally equivalent interleavings induce the same
happened-before relation, and vice versa.

\begin{lemma} \label{lemma:aux12}
  Let $S,S'$ be interleavings with the same initial pid. 
  Then, we have $S\approx S'$ iff 
  $\leadsto_S = \leadsto_{S'}$.
\end{lemma}
%
%
While interleavings might be closer to an actual execution,
it is often more convenient to have a higher-level representation,
one where all causally equivalent interleavings have the same
representation.
For this purpose, we introduce the notion of \emph{trace}
as a mapping from pids to sequences of actions.
Here, the key idea is to keep the actions of each process
separated. 

First, we introduce some notation. Let $\tau$ be a mapping
from pids to sequences of actions, which we denote
by a finite mapping of the form 
\[
[p_1\mapsto A_1;
\ldots; p_n\mapsto A_n]
\]
Given an interleaving
$S$, we let 
\[
\trace(S) = [p_1\mapsto \actions(p_1,S);
\ldots; p_n\mapsto\actions(p_n,S)]
\]
where $p_1,\ldots,p_n$
are the pids in $S$. 
We also let $\tau(p)$ denote 
the sequence of actions associated with process $p$ 
in $\tau$. Also, $\tau[p\mapsto A]$ denotes that
$\tau$ is an arbitrary mapping such that $\tau(p) = A$;
we use this notation either as a condition on $\tau$ or 
as a modification of $\tau$. 
We  also say that $(p\cons a)\in\tau$ if $a\in\tau(p)$.
Moreover, we say that $p_1\cons a_1$ \emph{precedes}
$p_2\cons a_2$ in $\tau$, in symbols
$(p_1\cons a_1)\prec_\tau (p_2\cons a_2)$,  
if $p_1=p_2$, $\tau(p_1) = A$,
and $a_1$ precedes $a_2$ in $A$; otherwise, the (partial) 
relation is not defined. 

\begin{definition}[trace] \label{def:trace}
  A \emph{trace} $\tau$ with initial pid $p_0$
  is a mapping from pids to sequences of actions if
  $\trace(S) = \tau$ for some interleaving $S$ with
  initial pid $p_0$.
\end{definition}
One could give a more direct definition of trace by mimicking the
conditions of an interleaving, but the above, indirect definition
is simpler. 

A trace represents a so-called
\emph{Mazurkiewicz trace}
\cite{Maz86}, i.e., it represents a \emph{partial order relation}
(using the terminology of model checking \cite{God97}),
where all linearizations of this partial order represent
causally equivalent interleavings. 
In particular, given a trace $\tau$, 
we let $\sched(\tau)$ denote the set of all
\emph{causally equivalent} linearizations of the events 
in $\tau$, which is formalized as follows:

\begin{definition} \label{def:sched}
  Let $\tau$ be a trace with initial pid $p_0$.
  We say that an interleaving $S$ with initial pid $p_0$
  is a linearization
  of $\tau$, in symbols $S\in\sched(\tau)$, if 
  $\trace(S) = \tau$.  
\end{definition}
The following property is a trivial consequence of our
definition of function $\sched$:

\begin{lemma} \label{lemma:aux3}
  Let $S,S'$ be interleavings with the same initial pid 
  and such that $\actions(p,S) = \actions(p,S')$
  for all pid $p$ in $S,S'$. Then, $\trace(S)=\trace(S')$.
\end{lemma}

\begin{proof}
  The proof is a direct consequence of the definition of
  function $\trace$, since only the relative actions of
  each process are recorded in a trace.
\end{proof}
%
%
The next result states that all the interleavings in
$\sched(\tau)$ are indeed causally equivalent:

\begin{theorem}
	Let $\tau$ be a trace. 
	Then, $S,S'\in\sched(\tau)$ implies $S\approx S'$.
\end{theorem}

\begin{proof}
  Let us consider two different interleavings $S,S'\in\sched(\tau)$.
  By the definition of function $\trace$ and 
  Definition~\ref{def:sched},
  $S$ and $S'$ have the same
  events and the same initial pid. Also, both
  interleavings have the same
  relative order for the actions of each process. 
  Moreover, by definition of interleaving 
  (Definition~\ref{def:interleaving}), we know that 
  all events $p\cons a$ of a process (but the initial one)
  must be preceded
  by an event $p'\cons \spawn(p)$, and that
  all receive events $p\cons \rec(\ell,cs)$ must be
  preceded by a corresponding send event
  $p'\cons \send(\ell,v,p)$. Therefore, the happened-before
  relation induced from $S$ and $S'$ must be the same and,
  thus, $S\approx S'$ by Lemma~\ref{lemma:aux12}. 
  \qed
\end{proof}
%
%
Trivially, all interleavings in $\sched(\tau)$ induce the same
happened-before relation (since they are causally equivalent). 
We also say that $\tau$
induces the same happened-before relation (i.e., $\leadsto_S$ for 
any $S\in\sched(\tau)$) and denote it
with $\leadsto_\tau$.

The following result is also relevant to conclude that a trace
represents \emph{all and only} the causally equivalent
interleavings.

\begin{theorem}
  Let $\tau$ be a trace and $S\in\sched(\tau)$ an interleaving. 
  Let $S'$ be an interleaving with $S'\not\in\sched(\tau)$.
  Then, $S\not\approx S'$.
\end{theorem}

\begin{proof}
  Assume that $S$ and $S'$ have the same
  events and that $\actions(p,S)=\actions(p,S')$ for
  all pid $p$ in $S,S'$
  (otherwise, the claim follows trivially).
  Let us proceed by contradiction. Assume that
  $S'\not\in\sched(\tau)$ and $S\approx S'$.
  Since $\actions(p,S) = \actions(p,S')$
  for all pid $p$ in $S,S'$, 
  we have $\trace(S) = \trace(S')$ by Lemma~\ref{lemma:aux3}.
  Thus, $S'\in\sched(\tau)$, which contradicts our assumption.
\qed
\end{proof}

\begin{example} \label{ex:running}
	Consider the following trace $\tau$, where we abbreviate
	$\send(\ell_i,v_i,p_i)$ as $s_i$
	and $\rec(\ell_i,cs_i)$ as $r_i$:
  \[
  \begin{array}{lllll}
  [ & \mathsf{p1} & \mapsto & \spawn(\mathsf{p3}),
      \spawn(\mathsf{p2}), \spawn(\mathsf{p4}),  
        \spawn(\mathsf{p5}),r_5,s_7; ~~~~
   \mathsf{p2}  \mapsto   s_2;\\
   & \mathsf{p3} & \mapsto  & r_1,s_3,r_2,r_4,s_5,r_6;~~~~
   \mathsf{p4} \mapsto  r_3;~~~~
    \mathsf{p5}   \mapsto  s_1,s_4,s_8 & ]\\
  \end{array}
  \]
  A possible execution following this trace is
  graphically depicted in 
  Figure~\ref{fig:running}, where spawn 
  actions have been omitted for clarity.
  Moreover, we assume that arrowheads represent
  the point in time where messages are delivered to
  the target process.
  Despite the simplicity of traces, we can extract some 
  interesting conclusions. 
  For example, if we assume that $\tau$ is the 
  trace of a terminating execution, we might conclude that
  messages $\ell_7$ and $\ell_8$ are
  \emph{orphan} messages (i.e., messages that are sent 
  but never received) since there are no corresponding events
  $r_7$ and $r_8$ in $\tau$. 
  A possible interleaving in $\sched(\tau)$ that follows the
  diagram in Figure~\ref{fig:running} is as follows:
    \[
    \begin{array}{l}
    \mathsf{p_1}\cons\spawn(p_3),~ 
    \mathsf{p_1}\cons\spawn(p_2),~ 
    \mathsf{p_1}\cons\spawn(p_4), ~
    \mathsf{p_1}\cons\spawn(p_5), \\
    \mathsf{p_5}\cons s_1,~
    \mathsf{p_3}\cons r_1,~
   \mathsf{p_2}\cons s_2,~
    \mathsf{p_3}\cons s_3,~
    \mathsf{p_3}\cons r_2,~
    \mathsf{p_5}\cons s_4,~
    \mathsf{p_3}\cons r_4,\\
    \mathsf{p_4}\cons r_3,~
    \mathsf{p_3}\cons s_5,~
    \mathsf{p_4}\cons s_6,~
    \mathsf{p_1}\cons r_5,~
    \mathsf{p_3}\cons r_6,~
    \mathsf{p_1}\cons s_7,~
    \mathsf{p_5}\cons s_8
    \end{array}
	\]
	By swapping, e.g., events
	$\mathsf{p_2}\cons s_2$ and 
    $\mathsf{p_3}\cons s_3$ we get
    another interleaving in $\sched(\tau)$,
    and so forth. Note that $\mathsf{p_2}\cons s_2$ and 
    $\mathsf{p_3}\cons s_3$ are independent since
    $\mathsf{p_3}\cons s_3 \prec_\tau 
    \mathsf{p_3}\cons r_2$.
	
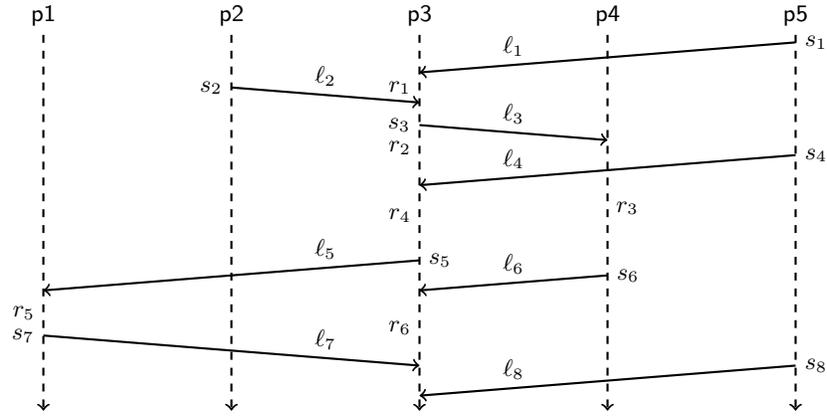
\begin{figure}[t]
\centering
\begin{tikzpicture}


\draw[->,dashed,thick] (0,5) node[above]{\sf p1} -- (0,0);
\draw[->,dashed,thick] (2.5,5) node[above]{\sf p2} -- (2.5,0);
\draw[->,dashed,thick] (5,5) node[above]{\sf p3} -- (5,0);
\draw[->,dashed,thick] (7.5,5) node[above]{\sf p4} -- (7.5,0);
\draw[->,dashed,thick] (10,5) node[above]{\sf p5} -- (10,0);

\draw[->, thick] (10,4.9) node[right]{$s_1$} -- (5,4.5) node[pos=0.75,above]{$\ell_1$} node[left]{}; 

\draw[->, thick] (2.5,4.3) node[left]{$s_2$} -- (5,4.1) node[midway,above]{$\ell_2$} node[right]{}; 

\draw[->, thick] (5,3.8) node[left]{$s_3$} -- (7.5,3.6) node[midway,above]{$\ell_3$} node[right]{}; 

\node at (5,4.3) [left] {$r_1$};

\node at (5,3.5) [left] {{$r_2$}};

\draw[->, thick] (10,3.4) node[right]{$s_4$} -- (5,3) node[pos=0.75,above]{$\ell_4$} node[left]{}; 

\node at (7.5,2.7) [right] {$r_3$};
\node at (5,2.6) [left] {$r_4$};

\draw[->, thick] (5,2) node[right]{$s_5$} -- (0,1.6) node[pos=0.25,above]{$\ell_5$} node[left]{}; 

\node at (0,1.3) [left] {$r_5$};

\draw[->, thick] (7.5,1.8) node[right]{$s_6$} -- (5,1.6) node[midway,above]{$\ell_6$} node[left]{}; 

\node at (5,1.1) [left] {$r_6$};

\draw[->, thick] (0,1) node[left]{$s_7$} -- (5,0.6) node[pos=0.75,above]{$\ell_7$} node[right]{}; 

\draw[->, thick] (10,0.6) node[right]{$s_8$} -- (5,0.2) node[pos=0.75,above]{$\ell_8$} node[left]{}; 
\end{tikzpicture}
\caption{Message-passing diagram. 
Processes ($\mathsf{pi}$, $i=1,\ldots,5$)  are represented
as vertical dashed arrows, where time flows from top to bottom. 
Message sending 
is represented by solid arrows labeled with a tag ($\ell_i$),
$i=1,\ldots,8$.
Note that all
events associated with a message $\ell_i$ have the same subscript $i$.}
 \label{fig:running}
\end{figure}
\end{example}
In the following, we assume that the actions of a process
are uniquely determined by the order of its receive
events (equivalently, by the order in which messages
are delivered to this process). 
To be more precise, given a sequence
of messages delivered to a given process, the actions of this
process are deterministic except for the choice of fresh identifiers
for the pids of spawned processes and 
the tags of sent messages, which has no impact
on the outcome of the execution. 
Therefore, if we have two executions
of a program where each process receives the same messages and
in the same order, and perform the same number of steps, then
the computations will be the same (identical, if we assume that
the same process identifiers and message tags are chosen). 

The following notion of \emph{subtrace} is essential to
characterize message races:

\begin{definition}[subtrace] \label{def:subtrace}
Given traces $\tau,\tau'$ with the same initial pid, 
we say that $\tau'$ is a \emph{subtrace} of $\tau$,
in symbols $\tau' \ll \tau$, iff for all pid $p$ in
$\tau,\tau'$ we have that the sequence $\tau'(p)$ is a prefix
of the sequence $\tau(p)$.
\end{definition}
Intuitively speaking, we can obtain a subtrace by deleting
the final actions of some processes. However, note that
actions cannot be arbitrarily removed since the resulting 
mapping must still be a trace (i.e., all linearizations
must still be interleavings according to 
Definition~\ref{def:interleaving}). For instance, this
prevents us from deleting the sending of a message
whose corresponding receive is not deleted.

Let us conclude this section 
with a \emph{declarative} notion of message race:

\begin{definition}[message race]
  Let $\tau$ be a trace with $\tau(p) = a_1,\ldots,a_n$
  and $a_i = \rec(\ell,cs)$, $1\leq i\leq n$. 
  %
  There exists a message race between $\ell$ and
  $\ell'$ in $\tau$ iff there is a subtrace $\tau'\ll\tau$
  such that $\tau'(p) = a_1,\ldots,a_{i-1}$ and
  $\tau'[p\mapsto a_1,\ldots,a_{i-1},\rec(\ell',cs)]$
  is a trace.
\end{definition}
Informally speaking, we have a message race whenever
we have an execution which is a prefix of the original one
up to the point where a different message is received.

Consider, e.g., the trace in Example~\ref{ex:running}.
If we assume that the value of message $\ell_4$
matches the constraints $cs_2$ of receive 
$r_2 = \rec(\ell_2,cs_2)$, i.e., $\match(v_4,cs_2)=\true$,
we have a race between  $\ell_2$ and $\ell_4$ 
since we have the following subtrace $\tau'$
\[
\begin{array}{lllll}
[ & \mathsf{p1} & \mapsto & \spawn(\mathsf{p3}),
    \spawn(\mathsf{p2}), \spawn(\mathsf{p4}),\spawn(\mathsf{p5}),\cancel{r_5},\cancel{s_7}; ~~~~
 \mathsf{p2}  \mapsto   s_2;\\
 & \mathsf{p3} & \mapsto  & r_1,s_3,\cancel{r_2},\cancel{r_4},\cancel{s_5},\cancel{r_6};~~~~
 \mathsf{p4} \mapsto  r_3;~~~~
  \mathsf{p5}   \mapsto  s_1,s_4,s_8 & ]\\
\end{array}
\]
and $\tau'[\mathsf{p3} \mapsto r_1,s_3,\rec(\ell_4,cs_2)]$ is a
trace according to Definition~\ref{def:trace}.

\section{Computing Message Races
and Race Variants} \label{sec:races}

In this section, we introduce constructive definitions for
computing message races and \emph{race variants} from a given 
execution trace. Intuitively speaking, once we identify a message
race, a race variant is a \emph{partial} trace that can be used
to drive the execution of a program so that a new interleaving
which is not causally equivalent to the previous
one is obtained.
Computing message races and race variants
are essential ingredients of a systematic 
state-space exploration tool.

First, we introduce the notion of \emph{race set} that,
given a trace $\tau$ and a message $\ell$ that has been
received in $\tau$, computes all the messages that race
with $\ell$ in $\tau$ for the same receive (if any).
It is worthwhile to note 
that race sets are defined on traces, i.e., message races
do not depend on a particular interleaving but on the class of 
causally equivalent interleavings represented by a trace.

\begin{definition}[race set] \label{def:race-set}
	Let $\tau$ be a  trace with
	$e_r =(p\cons\rec(\ell,cs))\in\tau$. 
	Consider a message $\ell'\neq\ell$ with 
	$e'_s =(p'\cons\send(\ell',v',p))\in\tau$ such that
	$\match(v',cs)=\true$. 
	We say that messages $\ell$ and $\ell'$ race for $e_r$ in $\tau$
	if 
   \begin{itemize}
	\item $e_r$ does not happen before $e'_s$, i.e., 
	$e_r\not\leadsto_\tau e'_s$;
	\item for all event $e''_s =(p'\cons \send(\ell'',v'',p))\in\tau$
	such that $e''_s \prec_\tau e'_s$ either $\match(v'',cs)=\false$
	or there exists an event $(p\cons \rec(\ell'',cs''))\in\tau$
	with $(p\cons \rec(\ell'',cs'')) \prec_\tau (p\cons \rec(\ell,cs))$.
	\end{itemize}	
	We let $\race_\tau(\ell)$ denote the set all messages that 
	race with $\ell$ in $\tau$.
\end{definition}
Intuitively speaking, the definition above requires the following
conditions for messages $\ell$ and $\ell'$ to race for a 
receive statement $e_r$:
\begin{enumerate}
\item The target of both messages must be the same ($p$)
 and their values should match the constraint $cs$ in $e_r$
(note that we already know that the value of 
message $\ell$ matches the 
constraint of $e_r$ since $\tau$ is a trace).

\item The original receive event, $e_r$, cannot happen before
the sending event $e'_s$ of message $\ell'$. 
Otherwise, we had a dependency and  removing
$e_r$ would prevent $e'_s$ to happen
(in a well-formed trace).

\item Finally, we should check that there are no other
messages sent by the same process (and to the same target)
that match the constraint $cs$ and have not been received
before $e_r$ (since, in this case, the first of such messages
would race with $\ell$ instead). 
\end{enumerate}
Given a trace $\tau$ and a receive event
$p\cons\rec(\ell,cs)\in\tau$, a naive algorithm for computing the
associated race set, $\race_\tau(\ell)$, can proceed as
follows:
\begin{itemize}
\item First, we identify the set of events of the form 
$p'\cons\send(\ell',v',p)$ in $\tau$ with $\ell' \neq\ell$,
i.e., all \emph{send} events where the target process is
$p$ and the message tag is different from $\ell$.
\item Now, we remove from this set each \emph{send} event
$p'\cons\send(\ell',v',p)$ where 
$p\cons\rec(\ell,cs)\leadsto_\tau p'\cons\send(\ell',v',p)$.
\item We also remove the events
$p'\cons\send(\ell',v',p)$ where $\match(v',cs) = \false$.
\item Finally, for each subset of \emph{send} events 
from the same process, we select (at most) one of them as follows. 
We check for each \emph{send} event (starting from the oldest one)
whether there is a corresponding \emph{receive} event in $p$ which
precedes $p\cons\rec(\ell,cs)$. The message tag of the
first \emph{send} event without a corresponding 
\emph{receive} (if any) belongs to the race set, and the 
remaining ones (from the same process) can be discarded.
\end{itemize}

\begin{example} \label{ex:running2}
   Consider again the trace $\tau$ from Example~\ref{ex:running}.
	Let us focus on the 
	second receive event of process $\mathsf{p3}$, denoted  by
	$r_2$. Here, we have five (other) messages with the same
	target ($\mathsf{p3}$): 	
	$\ell_1$, $\ell_4$, $\ell_6$, $\ell_7$ and $\ell_8$. 
	Let us further assume that
	the values of all messages match the constraint of
	$r_2$ except for message $\ell_4$. Let us analyze 
    each message separately:
	\begin{itemize}
	\item Message $\ell_1$ is excluded from the message race 
	since there exists a corresponding receive event, $r_1$,
	and $r_1\prec_\tau r_2$. Hence, $\ell_1\not\in\race_\tau(\ell_2)$.
	
	\item As for message $\ell_4$, we trivially have 
	$r_2\not\leadsto_\tau s_4$. Moreover, there is a previous
	event, $s_1$, from $\mathsf{p5}$ to $\mathsf{p3}$ but
	it has already been received ($r_1$). 
	However, we assumed that the value of message $\ell_4$
	does not match the constraints of $r_2$ and, thus,
	$\ell_4\not\in\race_\tau(\ell_2)$.
	
	\item Consider now message $\ell_6$. At first sight, 
	it may seem that there is
	a dependency between $r_2$ and the sending event $s_6$
	(since message $\ell_2$ was delivered before $s_3$).
	However, this is not the case since event $s_3$ happened
	before $r_2$ and, thus, $r_3\leadsto_\tau s_6$ but 
	$r_2\not\leadsto_\tau r_3$. Moreover, there are no previous send
	events in $\mathsf{p4}$ and, thus, $\ell_6\in\race_\tau(\ell_2)$.
	
	\item Regarding message $\ell_7$, we have 
	$r_2 \leadsto_\tau s_7$ since 
	$r_2 \leadsto_\tau s_5$, $s_5 \leadsto_\tau r_5$ and
	$r_5 \leadsto_\tau s_7$. Therefore, messages $\ell_2$
	and $\ell_7$ cannot race for $r_2$ and	
	$\ell_7\not\in\race_\tau(\ell_2)$.

	\item Finally, consider message $\ell_8$. 
   	Trivially, we have $r_2\not\leadsto_\tau s_8$. 
   	Now, we should check that all previous sent messages
   	($\ell_1$ and $\ell_4$) have been previously received or
   	do not match the constraints of $e_r$, which is the case.
   	Therefore, $\ell_8\in\race_\tau(\ell_2)$.
	\end{itemize}
	Hence, we have $\race_\tau(\ell_2) = \{\ell_6,\ell_8\}$.
\end{example}
As mentioned before, computing message races can be 
useful to identify alternative executions which are not
causally equivalent to the current one. Ideally, we want
to explore only one execution (interleaving) per  
equivalence class (trace).
For this purpose, we introduce the notion 
of \emph{race variant} which returns a (typically partial) trace,
as follows:

\begin{definition}[race variant]
Let $\tau[p\mapsto A\conc\rec(\ell,cs)\conc A']$
be a trace with $\ell'\in\race_\tau(\ell)$. 
The \emph{race variant} of $\tau$ w.r.t.\ $\ell$ and $\ell'$,
in symbols $\variant_\tau(\ell,\ell')$, is given by 
the  (possibly partial) trace
\[
\rdep(A',\tau[p\mapsto A\conc\rec(\ell',cs)])
\]
where the auxiliary function $\rdep$ is inductively defined as
follows:
\[
\begin{array}{l}
\rdep(A,\tau)  =\\
\hspace{2ex}\left\{\begin{array}{lll}
  \tau& \mbox{if}~A=\nil\\
  \rdep(A',\tau) & \mbox{if}~A=\rec(\ell,cs)\conc A'\\
  \rdep(A'\conc A'',\tau[p\mapsto\nil]) 
  & \mbox{if}~A=\spawn(p)\conc A',~\tau(p) = A''\\
  \rdep(A'\conc A^*,\tau[p\mapsto A'']) 
  & \mbox{if}~A=\send(\ell,v,p)\conc A',~\tau(p) = A''\conc\rec(\ell,cs)\conc A^*\\
  \rdep(A',\tau) 
  & \mbox{if}~A=\send(\ell,v,p)\conc A',~\rec(\ell,cs)\not\in\tau(p)\\  
\end{array}
\right.
\end{array}
\]
\end{definition}
Intutively speaking, 
$\variant_\tau(\ell,\ell')$ removes the original receive
action $\rec(\ell,cs)$ from $\tau$ as well as all the actions
that depend on this one (according to the happened-before
relation). Then, it adds $\rec(\ell',cs)$ in the position of the
original receive.

\begin{example}
  Consider again the execution trace $\tau$ from 
  Example~\ref{ex:running}, together with the associated race
  set computed in Example~\ref{ex:running2}: 
  $\race_\tau(\ell) = \{\ell_6,\ell_8\}$.
  Let us consider $\ell_6$. Here, the race variant 
  $\variant_\tau(\ell_2,\ell_6)$ is computed 
  from $\rdep((r_4,s_5,r_6),~
  \tau[\mathsf{p3}\mapsto r_1,s_3,\rec(\ell_6,cs)])$
  as follows:
  \[
  \begin{array}{ll}
  \rdep((r_4,s_5,r_6), &
  \tau[\mathsf{p3}\mapsto r_1,s_3,\rec(\ell_6,cs)])  \\
  = \rdep((s_5,r_6), &
  \tau[\mathsf{p3}\mapsto r_1,s_3,\rec(\ell_6,cs)])  \\
  = \rdep((r_6,s_7), &
  \tau[\mathsf{p1}\mapsto \spawn(\mathsf{p3}),
      \spawn(\mathsf{p2}), \spawn(\mathsf{p4}),  \spawn(\mathsf{p5});\\
  & ~~\:\mathsf{p3}\mapsto r_1,s_3,\rec(\ell_6,cs)])  \\
  = \rdep((s_7), &
  \tau[\mathsf{p1}\mapsto \spawn(\mathsf{p3}),
      \spawn(\mathsf{p2}), \spawn(\mathsf{p4}),  \spawn(\mathsf{p5});\\
  & ~~\:\mathsf{p3}\mapsto r_1,s_3,\rec(\ell_6,cs)])  \\
  = \rdep(\nil, &
  \tau[\mathsf{p1}\mapsto \spawn(\mathsf{p3}),
      \spawn(\mathsf{p2}), \spawn(\mathsf{p4}),  \spawn(\mathsf{p5});\\
  & ~~\:\mathsf{p3}\mapsto r_1,s_3,\rec(\ell_6,cs)])  \\
  \end{array}
  \]
  Therefore, the computed race variant  $\tau'$ is as follows:
  \[
  \begin{array}{lllll}
  [ & \mathsf{p1} & \mapsto & \spawn(\mathsf{p3}),
      \spawn(\mathsf{p2}), \spawn(\mathsf{p4}),  \spawn(\mathsf{p5}); ~~~~
   \mathsf{p2}  \mapsto   s_2;\\
   & \mathsf{p3} & \mapsto  & r_1,s_3,\rec(\ell_6,cs);~~~~
   \mathsf{p4} \mapsto  r_3;~~~~
    \mathsf{p5}   \mapsto  s_1,s_4,s_8 & ]\\
  \end{array}
  \]
\end{example}
In the following, given traces $\tau,\tau'$, if $\tau$ is
a subtrace of $\tau'$, i.e., $\tau\ll\tau'$, we also say that
$\tau'$ \emph{extends} $\tau$.
Let us consider a trace $\tau$ and one of its race variants
$\tau'$. The next result states that there is no interleaving
in $\sched(\tau'')$ that is causally equivalent to any
interleaving of $\sched(\tau)$ for all traces $\tau''$
that extend the race variant $\tau'$.
This is an easy but essential property to guarantee the 
optimality in the number of variants considered
by a state-space exploration algorithm.

\begin{theorem} \label{th:race-variant}
	Let $\tau$ be a trace with $e_r = (p\cons\rec(\ell,cs))\in\tau$
	and $\ell'\in\race_\tau(\ell)$. 
	Let $\tau' = \variant_\tau(\ell,\ell')$
	be a race variant. Then, for all trace $\tau''$ that
	extends $\tau'$ and for all interleavings
	$S\in\sched(\tau)$ and $S''\in\sched(\tau'')$,
	we have $S\not\approx S''$.
\end{theorem}

\begin{proof}
  Consider first a (possibly partial)
  trace $\tau_1$ obtained from 
  $\rdep(A',\tau[p\mapsto A\conc\rec(\ell,cs)])$, i.e., $\tau_1$
  is equal to the race variant except
  for the fact that we have not changed yet the considered
  receive event. Then, it is easy to see that $\tau_1$ is
  a subtrace of $\tau$, $\tau_1\ll\tau$, since $\rdep$ just follows the 
  happened-before relation in order to consistently remove
  all dependences of $e_r$. Note that $\tau_1$ and $\tau'$
  only differ in the receive event ($\rec(\ell,cs)$ in $\tau_1$
  and $\rec(\ell',cs)$ in $\tau'$).
  Trivially, for all 
  interleavings $S_1\in\sched(\tau_1)$ and $S'\in\sched(\tau')$,
  we have $S\not\approx S'$ since the receive events 
  $\rec(\ell,cs)$ and $\rec(\ell',cs)$ can only happen in
  one of the interleavings but not in both of them. Moreover,
  for all trace $\tau''$ that extends $\tau'$, and for all
  interleavings $S''\in\sched(\tau'')$ and $S\in\sched(\tau)$,
  we have $S''\not\approx S$ since they will always differ
  in the receive events above. \qed
\end{proof}
The definitions of message race and race variant can be used
as the kernel of a state-space exploration technique 
that proceeds as follows:
\begin{enumerate}
\item First, a random execution of the program is considered,
together with its associated trace.
\item This trace is used to compute message races (if any) 
as well as the corresponding race variants. 
\item Then, each computed 
race variant is used to drive the execution of the
program up to a given point, then continuing the execution
nondeterministically according to the standard semantics.
We gather the traces of these executions and the process
starts again until all possible executions have been explored.
\end{enumerate}
A formalization of such an algorithm can be found in
the context of reachability testing \cite{LC06} using
SYN-sequences instead of traces. 

On the other hand, 
the \emph{prefix-based tracing} technique for Erlang 
introduced in \cite{GV21lopstr} 
could be useful to instrument programs
with a (possibly partial) trace so that their execution 
follows this trace and, then, continues nondeterministically,
eventually producing a trace of the complete execution
(point 3 above).
The notion of trace in \cite{GV21lopstr}
is different to our notion of trace, though: message
delivery is explicit and traces do not include message values
nor receive constraints. Nevertheless, 
adapting their developments to
our traces would not be difficult.

The definitions of message race and race variant could also
be useful in the context of causal-consistent
replay debugging \cite{LPV19,LPV21}. First, we note that
our traces could be straightforwardly 
used for replay debugging since they contain
strictly more information than the logs of \cite{LPV19,LPV21}.
However, in contrast to the original logs, our traces would
allow the replay debugger \cauder\ \cite{cauder} 
to also show the message races in a particular
execution, and then let the user to replay any selected race variant,
thus improving the functionality of the debugger.
Some ongoing work along
these lines can be found in \cite{GV21arxiv}.
However, the traces considered in \cite{GV21arxiv} 
are similar to those in \cite{GV21lopstr} 
(i.e., they have explicit events for message delivery and 
skip message values and receive constraints).
As a consequence, the races considered in
\cite{GV21arxiv} are only \emph{potential} races since 
there are no guarantees that message
values in these races actually match the corresponding
receive constraints.
Nevertheless, an extension of \cauder\ using our
traces and the associated definitions of message race
and race variant could be defined following a similar scheme.


\section{Discussion and Future Work}\label{sec:relwork}

The closest approach to our notion of trace are the
\emph{logs} of \cite{LPV19,LPV21}, which where
introduced in the context of causal-consistent \emph{replay} 
debugging for a message-passing
concurrent language. In this work, we have
extended the notion of log with enough information so that
message races can be computed.
Indeed, this work stemmed from the idea of 
improving causal-consistent 
replay debugging \cite{LPV19,LPV21} with the computation of 
message races, since this information might be useful for the user 
in order to explore alternative execution paths. A first implementation
in this direction is described in \cite{GV21arxiv}, although
the traces are slightly different, as discussed above.

Another close approach is that of \emph{reachability testing},
originally introduced in \cite{HTH94} in the context of 
multithreaded programs that perform read/write operations.
This approach was then extended to message-passing 
programs in \cite{Tai97,LT02} and later improved and
generalized in \cite{LC06}.\footnote{\cite{NM95} also deals with
message-passing concurrent programs, but only
\emph{blocking} send and receive statements are considered.}
 The notion of \emph{SYN-sequence}
in reachability testing (and, to some extend, the
\emph{program executions} of \cite{CL95}) share some
similarities with our traces since both
represent a partial order with the actions performed
by a number of processes running concurrently (i.e., they basically
denote a \emph{Mazurkiewicz trace} \cite{Maz86}).
Nevertheless, our traces
are tailored to a language with selective receives by
adding message values and receive constraints 
(\cite{LC06}, in contrast, considers
different ports for receive statements). 
Moreover, to the best of our knowledge, these works
have not considered a language where
processes can be dynamically spawned, as we do.

Both reachability testing and our approach share some similarities
with so-called \emph{stateless model checking}  \cite{God97}.
The main difference, though, is that stateless model checking works 
with interleavings.
Then, since many interleavings may boil down to the same
Mazurkiewicz trace, \emph{dynamic partial order reduction} (DPOR)
techniques are introduced (see, e.g., \cite{FG05,AAJS17}).
Intuitively speaking, DPOR techniques aim at producing only one
interleaving per Mazurkiewicz trace. 
Computing message races is 
more natural in our context thanks to the use of
traces, since
DPOR techniques are not needed.
Concuerror \cite{CGS13} implements stateless model checking
for Erlang \cite{AAJS17,AS17}, and has been recently 
extended to also consider
\emph{observational equivalence} \cite{AJLS18}, thus achieving
a similar result as our technique regarding the computation
of message races,
despite the fact that the techniques are rather different
(using traces vs using interleavings + DPOR).

Another, related approach is the detection of race 
conditions for Erlang programs presented in \cite{CS10}. 
However, the author focuses on data races (that may occur when
using some shared-memory built-in operators of the language) 
rather than message races.
Moreover, the detection is based on a \emph{static} analysis, while
we consider a \emph{dynamic} approach to computing message
races.


To conclude, we have introduced appropriate
notions of interleaving and trace  
that are useful to represent
concurrent executions in a message-passing concurrent
language with dynamic process spawning and
selective receives. In particular, our notion of
trace is essentially equivalent to 
a Mazurkiewicz trace, thus allowing us to represent
all causally equivalent interleavings in a compact way. 
Despite the simplicity of traces,
they contain enough information to analyze some common
error symptoms (e.g., orphan messages) and to compute
message races, which can then give
rise to alternative executions (specified by so-called
race variants, i.e., partial traces).

As for future work, we will consider the computation of
message races from \emph{incomplete}
traces, since it is not uncommon that concurrent programs
are executed in an endless loop and, thus, the associated
traces are in principle infinite. We also plan to extend the
traces with more events (like message
\emph{deliver} and process \emph{exit}) so that they can
be used to detect more types of error symptoms 
(like process deadlocks and lost or delayed messages).

Finally, another interesting line of research involves formalizing
and implementing
an extension of the causal-consistent replay
debugger \cauder\ \cite{cauder} for Erlang
in order to also
show message races
(our original motivation for this work). 
A preliminary approach along these lines
can be found in \cite{GV21arxiv}, though the considered
traces are slightly different, as mentioned above. 
In this context, we also plan to analyze
efficiency issues and investigate the definition of 
efficient algorithms for computing race sets.


\subsubsection*{Acknowledgements.}

The author would like to thank 
Juan Jos\'e Gonz\'alez-Abril 
for his useful remarks 
on a preliminary version of this paper.
I would also like to thank the anonymous 
reviewers for their suggestions to improve this work.

\bibliographystyle{splncs04}

\end{document}